\documentclass{article}

\usepackage{arxiv}

\usepackage[utf8]{inputenc} 
\usepackage[T1]{fontenc}    
\usepackage{hyperref}       
\usepackage{url}            
\usepackage{booktabs}       
\usepackage{amsfonts}       
\usepackage{nicefrac}       
\usepackage{microtype}      
\usepackage{lipsum}		
\usepackage{graphicx}

\graphicspath{ {./figures/} }	
\usepackage[normalem]{ulem}	
\usepackage{cite}			
\usepackage{subcaption}		

\title{AI Chiller: An Open IoT Cloud Based Machine Learning Framework for the Energy Saving of Building HVAC System via Big Data Analytics on the Fusion of BMS and Environmental Data}

\hypersetup{
	pdfborder={0 0 0 [3 3]}
}

\author{ \href{https://orcid.org/0000-0002-5700-4542}{\includegraphics[scale=0.06]{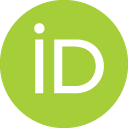}\hspace{1mm}Yong Yu}\thanks{Dr. Yong Yu is a Data Scientist at Siemens Ltd. Hong Kong. This work was done during his employment with Siemens.}\\
	Siemens Advanta Solutions \\
	New Territories, Hong Kong \\
	\texttt{advantasolutions.hk@siemens.com}\thanks{Ms. Jimalyn Yao is a Product Manager, Digital Solutions Manager at Siemens Ltd. Hong Kong, please contract her for any enquiries regarding this paper via \texttt{advantasolutions.hk@siemens.com}.} \\
}

\hypersetup{
pdftitle={AI Chiller: A Open IoT Cloud Based Machine Learning Framework for the Energy Saving of Building HVAC System via Big Data Analytics on the Fusion of BMS and Environmental Data},
pdfsubject={cs.AI, cs.LG, cs.NE, cs.SY, eess.SY}, 
pdfauthor={Jeremy ~Yu},
pdfkeywords={AI, Machine Learning},
}

\begin{document}
\maketitle
\hypersetup{
	pdfborder={0 0 0.5 [3 3]}
}
\begin{abstract}
	Energy saving and carbon emission reduction in buildings is one of the key
	measures in combating climate change. Heating, Ventilation, and Air
	Conditioning (HVAC) system account for the majority of the energy
	consumption in the built environment, and among which, the chiller plant
	constitutes the top portion. The optimization of chiller system power
	consumption had been extensively studied in the mechanical engineering and
	building service domains. Many works employ physical models from the domain
	knowledge. With the advance of big data and AI, the adoption of machine
	learning into the optimization problems becomes popular. Although many
	research works and projects turn to this direction for energy saving, the
	application into the optimization problem remains a challenging task. This
	work is targeted to outline a framework for such problems on how the energy
	saving should be benchmarked, if holistic or individually modeling should
	be used, how the optimization is to be conducted, why data pattern
	augmentation at the initial deployment is a must, why the	gradually
	increasing changes strategy must be used. Results of analysis on historical
	data and empirical experiment on live data are presented.
\end{abstract}


\section{Introduction}

\begin{figure}
	\centering
	\includegraphics[scale=0.9]{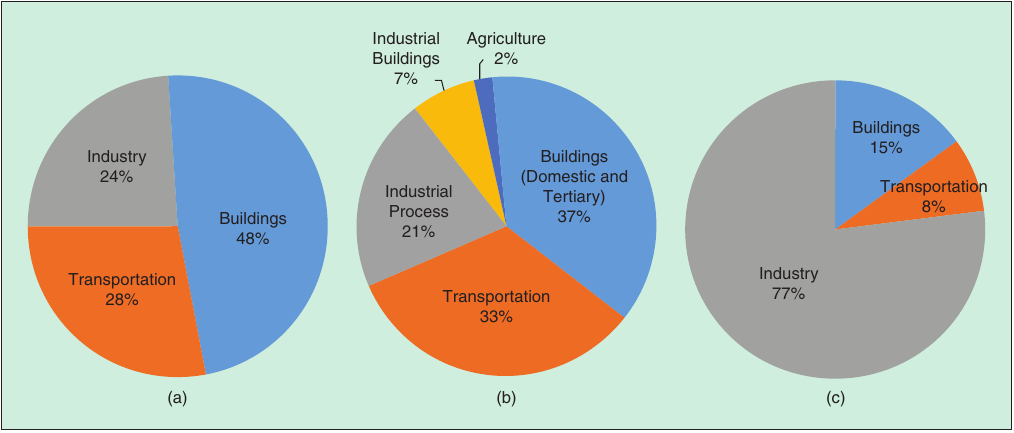}
	\caption{The energy consumption by sector for (a) the United States, (b) the European Union, and (c) China\cite{Manic_2016}}
	\label{fig:1}
\end{figure}

To combat climate change, the Paris Agreement aims to strengthen the global response by keeping global temperature rise this century well below 2 degrees Celsius above pre-industrial levels \cite{2015}. The United Nations Environment Program estimates that 40\% of global energy is used in buildings(Figure~\ref{fig:1})\cite{SBCI2015}. The residential and commercial buildings consume approximately 60\% of the world’s electricity\cite{Manic_2016,Laustsen2008,Weng_2012}. Because of the high energy consumption, the building sector has also been shown to provide the greatest potential for delivering significant cuts in emissions and cost globally\cite{SBCI2015}, and such potential has been projected to increase in the future\cite{Weng_2012, P_rez_Lombard_2008}. For metropolis like Hong Kong, 66.5\% of greenhouse gas emissions came from electricity in 2016, and Hong Kong committed to reduce its absolute carbon emissions by 26-36 per cent by 2030 from 2005 levels\cite{Poon2018}. Centralized HVAC plants are widely used in commercial buildings, and many are using water-cooled chiller systems during the last decades in substitution of the air-cooled chiller systems, which had increased the cooling efficiency significantly, and consumes up to 20\% less electricity\cite{EHK2020}. When new installation to new buildings or major retrofit to existing buildings to the water-cooled chiller systems exhibited remarkable savings during the last two decades, more potentials of savings are necessary to strive for more power savings.

When hardware installation and upgrade have been widely adopted, the industry turns to the optimization on the operations of the HVAC systems in search of more power savings. In deed there had been many research works on the optimization techniques of HVAC systems, especially chiller plants \cite{Braun1990,Wei_2014,Chow_2002}. In the fields of mechanical engineering and building service engineering, the chiller plant optimization methodologies usually rely on the physical models of the HVAC systems, and such models would have to be based on many assumptions on the system and equipment running conditions, which may or may not hold true for the studied target systems. Especially,  physical equipment and systems often vary due to the different installation conditions, extrinsic parameters, and the operation conditions often deviate across the lifetime of the equipment.

Modern buildings often come with Building Automation Systems (BAS), or Building Control Systems (BCS), typically consist of building (energy) management systems (BMS/BEMS), which control HVAC primary components such as air handling units (AHUs), chillers, and heating elements. As a major effort for the automation of building operations, such BMSs collect abundant data from the components they monitor and control, but they stick to the purpose of automation, not much value-added analysis and application on the data collected.  For the optimization of chiller systems to minimize the building energy consumption, a new generation of BMS which is armed with data driven models to achieve the energy saving by controlling and scheduling the components in the buildings is required.

In addition to data from BMS systems, with the advances in the IoT sensor technologies, it becomes convenient and economical to install IoT sensors to collect live data and transmit back via wireless channel. This makes it possible to collect whatever data required for the purpose of chiller plant optimization.  On the other hand, the booming of AI/Machine Learning (ML) power and technologies enables the capability of in-depth analytics on the data collected. Hence advanced and flexible optimization algorithms and systems becomes feasible by applying AI/ML models on the historical and live data from BMS and/or IoT sensors.

The building HVAC system in itself is a complex system which consists of many interconnected electrical and mechanical machines, and the system is affected by many extrinsic and intrinsic factors that make the system extremely hard to model. But the prevalent adoption of IoT and big data based infrastructure and the maturity of ML, especially Deep Learning (DL) algorithms makes the optimization of such system possible.

There are two main directions on the power savings of the HVAC systems, one is to reduce the cooling production directly, the other is to reduce the power consumption when at the same time maintain a given cooling production profile. In deed, in either way, one needs to set the appropriate cooling production targets or benchmarks, which could be a fixed value, a predefined target, or some other cooling profile calibred/forecasted by historical patterns, real-time occupancy, demand etc.. The former direction is to directly lower the production, but in many circumstances, one has to guarantee some service level or fulfil some given demand. Although this is an easy direction to go on the optimization aspect, effort on identifying the actual demand to be fulfilled must be made. The latter direction is to setup a cooling target, or cooling target profile, and to optimize the HVAC system to achieve the optimal power consumptions, while achieving the target or target profile at the same time.

We go for the data-driven optimization approach, in stead of making many assumptions and rely highly on over simplified theoretical models, we take advantage of the abundance of data and train models based on the big data, which take various system extrinsic and intrinsic factors into consideration, hence is highly adaptive to many circumstances like aging devices, deteriorating working conditions, etc.. On one hand, we recommend to use AI/ML algorithms, which in nature the appropriate companion to big data problems, in the modelling of big data, and we recommend to model the whole HVAC system as a whole under such conditions. On the other hands, we recognize that in some situations, when data are limited and time is constrained, traditional algorithms/models combined with domain expert knowledge might be the right approach for the startup of the whole implementation of such optimization problems, and also sometimes modelling individual components of the system might be the better choice.

In this work, we first outline a framework towards the energy savings on the building HVAC system, we then elaborate on the directions and  present an empirical experiment on the implementation of the later direction. Finally we draw the conclusion.

\section{Literature Review}
\label{sec:Literature Review}
In this section, we review the previous works on the energy saving and optimization problems on the HVAC systems.

HVAC systems had been extensively studied in the mechanical engineering and building service domains, and the optimization of the systems to achieve energy consumption/cost savings had been one of the key concentrations. In \cite{Weaver1975}, Weaver presented several approaches to the energy savings in the HVAC area with the use of closed-loop computer control and optimization. In \cite{Doukas_2007}, Doukas et al. studied a decision support model using rule sets based on a typical building energy management system,  which incorporated the requirements in both guaranteeing of the desirable levels of living quality in all building's rooms and the necessity for energy savings. In \cite{OLSON_1990}, a mathematical programming approach for determining which available chiller plant equipment to use to meet a cooling load as well as the best operating temperatures for the water flows throughout the system was presented.
There was work specifically for the optimizing variable-speed pumps of indirect water-cooled chilling systems \cite{Wang_2001}, they use online control strategy and find the optimal control by adjusting the pressure set-point according to the estimated derivative of the total power with respect to pressure. Ben-Nakhi and Mahmoud \cite{Ben_Nakhi_2004} designed a General regression neural networks (GRNN) algorithm to predict cooling load for buildings, and further used the algorithm to optimize the HVAC system in office buildings. Their results showed a properly designed NN is a powerful instrument for optimizing thermal energy storage in buildings based only on external temperature records. Wei et al.\cite{Wei_2014} deemed that the chiller plant model derived from data-driven approach is a nonlinear and non-convex optimization problem, and derived a two-level intelligent algorithm to solve the model aiming at minimizing the total cost of the chilled water plant, a simulation case was conducted and the corresponding results were discussed.

With the popularity of big data and deep learning (DL) during the 2010s\cite{Bengio2007,Kersting2018,SAS2018}, DL had been used in much broader scenarios, and applying DL in the data centers to saving energy and money was one of the natural expansions. E.g., DeepMind \cite{Gao2014, Evans2016} optimized the power for cooling service in Google's data centers, based on the patterns of energy consumption linked to running status of the machines in the data center. Siemens \cite{Judge2018, Kreutzer2018} also offered an AI-powered thermal optimization service for data centers. Inspired by these applications, more use cases on the power savings of the commercial buildings were witnessed \cite{Zheng:2018:DDC:3208903.3208913,Vishwanath2019,Vishwanath2019a, Vu:2017:DDC:3132847.3132860}.
In \cite{Li_2020} reinforcement learning, Li et al. proposed optimizing the data center cooling control via the emerging deep reinforcement learning (DRL) framework, which provided an end-to-end cooling control algorithm.

\section{A Framework for the Energy Optimization of the HVAC systems}
\begin{figure}
	\centering
	\includegraphics[scale=0.5]{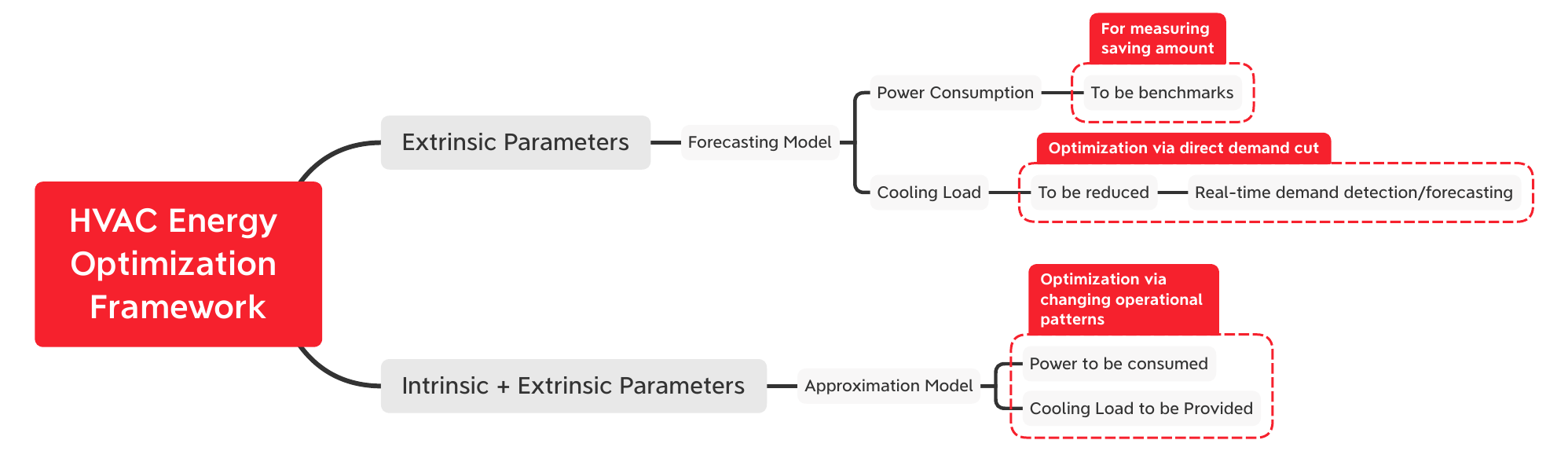}
	\caption{A Framework for the Energy Optimization of the HVAC systems}
	\label{fig:2}
\end{figure}

Many of the previous works focused on the optimization part of the HVAC power saving problems, but put not much effort on the measurement of the savings. Some of them use averaged previous power consumptions as the benchmark, others use averaged same period (e.g. same month) last year(s). Either way, these approaches introduced significant bias. Our analysis reveals that the extrinsic weather conditions plays the most important role affecting the building HVAC demands, and hence the power consumptions. Therefore using just past power consumptions as the benchmark without considering the corresponding weather conditions is subject to inaccuracy. Accordingly, in our framework, we propose to setup the appropriate benchmark via building a dedicated model, which forecast the cooling load demand and the corresponding power consumption, as shown in Figure~\ref{fig:2}.

We then outline that the first choice for the optimization is to reduce the cooling demand directly, or change the cooling profile such as pre-cooling. The other choice is to keep the original, or previous cooling load (corresponding to the weather conditions), which search the possible way of saving power via changing the operational patterns. Model(s) is needed to simulate the performance of the HVAC system, which then can generate searching space with both intrinsic parameters and extrinsic parameters.

In the following sections, we introduce an empirical experiment on the optimization of energy user of a Chiller Plant in a office building.

\section{Benchmarking Methodology}
\begin{figure}
	\centering
	\includegraphics[scale=0.9]{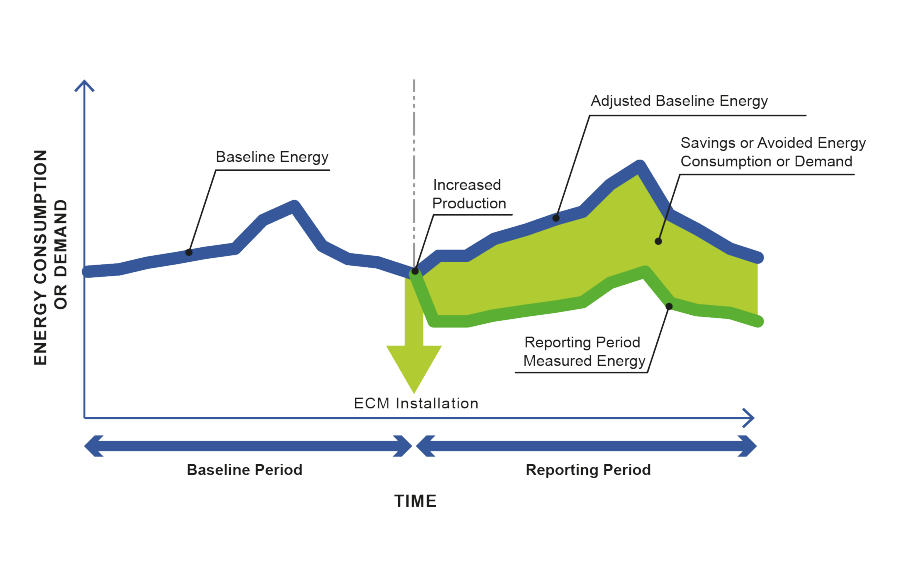}
	\caption{Benchmarking of Savings\cite{EVO1000012012}}
	\label{fig:3}
\end{figure}

It is usually very hard, if not impossible, to directly measure the savings on energy, water or demand \cite{EVO1000012012}, since savings represent the
absence of energy, water use or demand. Instead, savings often have to be determined by comparing
measured use or demand before and after the implementation of a program, making suitable
adjustments for changes in conditions. One can never obtain both actual measurement at the same.
Figure~\ref{fig:3} shows the energy-usage history of
a HVAC device before and after the deployment of an energy conservation measure (ECM). At about the time of installation, the demand for the plant production also increased.
The energy effect must be separated from that of the increased demand in order to properly document the savings of the ECM. The “baseline energy” use the patterns before the ECM installation to determine the relationship between energy use and the demand on production.
After the ECM installation, such baseline was used to estimate how much energy the plant would have used each month if there had been no ECM (called the “adjusted-baseline
energy”). The saving, or ‘avoided energy use’ is the difference between the adjusted-baseline
energy and the energy that was actually metered during the reporting period\cite{EVO1000012012}.
Without the adjustment for the change in production, the difference between baseline energy
and reporting period energy would have been much lower (or higher depending on the actual conditions). This will introduce significant bias on the calculation of savings.
It is necessary to clearly separate the energy effects of a savings program from the effects of other simultaneous changes affecting the energy using systems. The comparison of before and after
energy use or demand should be made on a consistent basis. Hence, in \cite{EVO1000012012}, an "Adjustments" term is introduced to re-state the use or demand of the baseline and reporting periods under a common set of conditions. This adjustments term
distinguishes proper savings reports from a simple comparison of cost or usage before and after
implementation of an ECM. Simple comparisons of utility costs
without such adjustments report only cost changes and fail to report the true performance of a
project. To properly report “savings,” adjustments must account for the differences in conditions
between the baseline and reporting periods.
The baseline in an existing facility project is usually the performance of the facility or system
prior to modification. This baseline physically exists and can be measured before changes are
implemented. In new construction, the baseline is usually hypothetical and defined based on
code, regulation, common practice or documented performance of similar facilities. In either
case, the baseline model must be capable of accommodating changes in operating parameters
and conditions so “adjustments” can be made.

\begin{figure}
	\centering
	\begin{subfigure}{0.31\textwidth}
		\includegraphics[width=\linewidth]{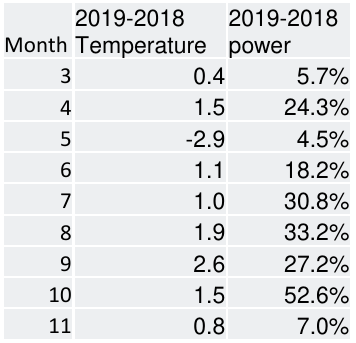}
		\caption{Monthly averaged temperature vs. change of power usage} \label{fig:4a}
	\end{subfigure}%
	\hspace*{\fill}   
	\begin{subfigure}{0.31\textwidth}
    	\includegraphics[width=\linewidth]{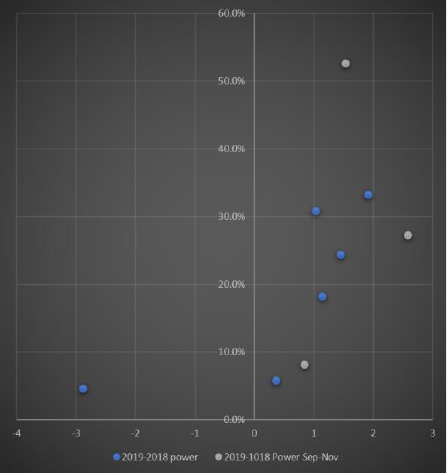}
    	\caption{Plot of temperature vs. power} \label{fig:4b}
  	\end{subfigure}%
  	\hspace*{\fill}   
  	\begin{subfigure}{0.31\textwidth}
    	\includegraphics[width=\linewidth]{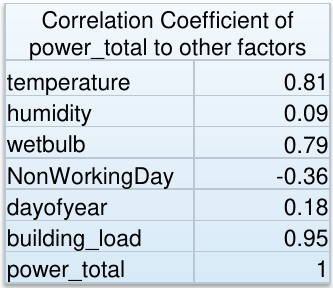}
    	\caption{Correlation between extrinsic parameters and power usage} \label{fig:4c}
  	\end{subfigure}
	\caption{Benchmarking of Savings}
	\label{fig:4}
\end{figure}

Historical data from the weather station and building BMS data are used in the analysis the relationship between extrinsic parameters and the power usage of the building chiller plant. Our results reveal that there is strong correlation between the weather temperature and the power usage, as shown in Figure~\ref{fig:4}. A deep learning model, specifically the RNN-LSTM model, is employed in the forecasting of power usage and cooling load demand. We achieve Mean Absolute Percentage Error (MAPE) 16.6\% ±10\% (95\% confidence level, same for MAPE hereafter) in the forecasting for power usage, and MAPE 14.0\% ±5.8\% for cooling load demand. The results illustrated that there are strong correlations between weather conditions and the building cooling load and power usage, and therefore one can forecast the the load and power from weather conditions quite accurately. Though weather plays the prominent role in affecting the cooling and power, the MAPEs all greater than 10\% also exhibit that there must be other factors which are not as important, but still considerable. We regard the actual building occupancy as one of such extrinsic factors, and we leave it for future study.

\section{Modelling Methodology}

\subsection{Holistic Model over Chiller Plant}
\label{sec:method-Holistic}
There two possible methods in modeling the Chiller Plant system. One is to model the whole system as a whole, the other is to model each of the major devices individually, and combine such sub-models together at the end \ref{Vu:2017:DDC:3132847.3132860}. The former one comes with simplicity that only the extrinsic parameters and outputs will be considered, and all the intermediate parameters are ignored. The latter, on the contrary, requires intermediate parameters that are related to those individual devices, hence require more data. Though the latter one had the benefit of easy to start and easy to incorporate domain expert knowledge, the higher data requirement also makes it hard to generalize.

In our test case, our BMS data lack of some important intermediate parameters like flow rate of water pipes, hence we adopt the holistic model.



\subsection{Optimization Algorithm}
\label{sec:others}

The holistic model of the chiller plant with 5 chillers, 12 pumps and 4 cooling towers consists of almost 80 input parameters and 2 outputs, such very high dimensional optimization problem is hard to be solved by conventional operations research algorithms. We employ the meta-heuristic algorithms in finding the optimal solutions. Particle Swarm Optimization (PSO) is found to converge faster, but the results are unstable. The Genetic Algorithm (GA) is slower, but it manage to finish the optimization within the time frame of hours, and converge stably. Hence we adopt the GA as our optimization algorithm.

\section{Results and Discussion}

\begin{figure}
	\centering
	\includegraphics[scale=0.8]{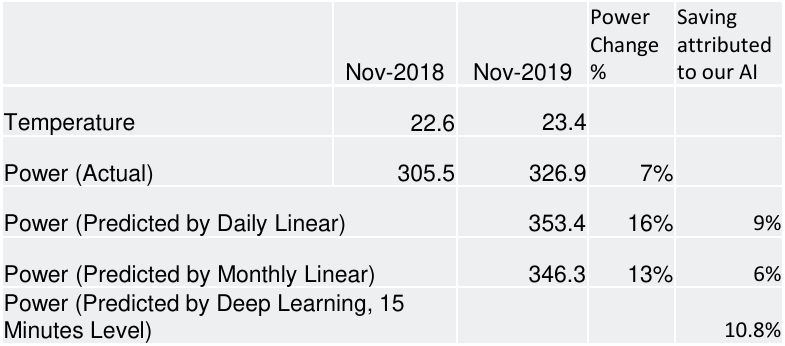}
	\caption{Results on Savings}
	\label{fig:5}
\end{figure}

\begin{figure}
	\centering
		\begin{subfigure}{0.5\textwidth}
			\includegraphics[width=\linewidth]{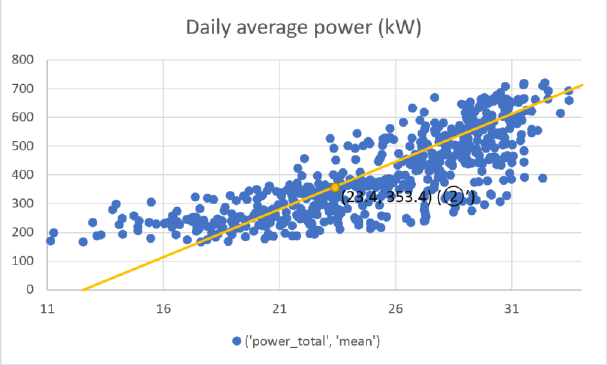}
			\caption{Daily averaged temperature vs. power usage} \label{fig:6a}
		\end{subfigure}
		\begin{subfigure}{0.4\textwidth}
			\includegraphics[width=\linewidth]{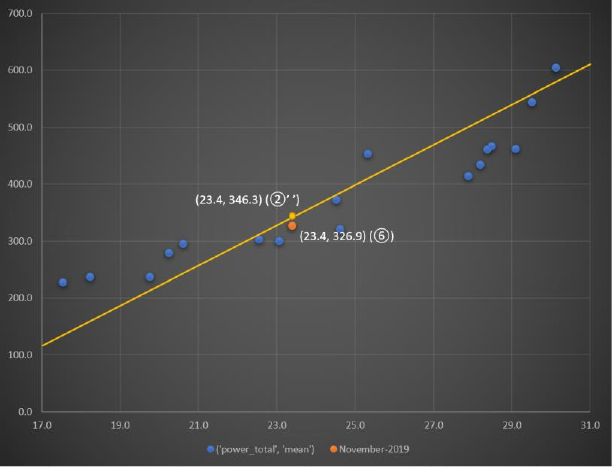}
			\caption{Monthly averaged temperature vs. power usage}\label{fig:6b}
		\end{subfigure}
	\caption{Savings by Linear Models}
	\label{fig:6}
\end{figure}

Our whole solution is implemented based on 18 months historical data from Mar 2018 to Aug 2019 and deployed to the studied building and with our recommendations into actual operations. Results are collected for two months (Oct 2019-Nov 2019), and the final results are shown in Figure~\ref{fig:5} and Figure~\ref{fig:6}.

\subsection{The necessity of data augmentation}
During the early deployment of our solution, we observed forecasting performance drop of the chiller plant simulation model. Such phenomena
was also witnessed in \cite{Vu:2017:DDC:3132847.3132860}. When the model is trained on the historical data, which in our case, are collected from the BMS that is running under the old operational practices and contain limited and only old patterns, it learns only old patterns. When it is deployed, and our recommendations are put in operations, new patterns emerge, the model performance then dropped. Therefore, right after the solution is deployed, a time period for the data (pattern) augmentation is required. The model must be retrained with data colleted from this period. Subsequently, the performance is only acceptable after the retraining process is complete.

\subsection{Savings from deep learning benchmarks}
In our solution, we target at providing the "same" cooling profile. Such cooling profile comes from the forecasting model which takes the weather temperature and humidity as the input parameters. Our optimization algorithms search within the feature space generated from a second model which simulates the chiller plant performance. Our final results of the last month show that 10.8\% power savings was achieved. This is obtained from comparing the actual power usage to the forecasted cooling profile in 15 minute intervals.

\subsection{Savings from linear model benchmarks}
As a comparison, we also conducted two linear analyses. One of them used the daily averaged temperature and power usage to forecast the daily power usage from daily average temperature. The other used the monthly averaged daily temperature and daily power usage to do the same forecasting. 9\% and 6\% savings are observed, which cross-validated practical savings achieved.

\subsection{Future Directions}
In this work, we outlined a framework for the energy savings on the building HVAC system, and presented an empirical experiment on the chiller plant system, with around 10\% savings achieved. As we have also discussed in Section \ref{sec:method-Holistic}, it is worth to try power savings on the air-side devices, i.e., the Air Handling Unit (AHU). Both methods can constitute a full picture for the power savings on HVAC system.

Another direction is to make a bigger picture of the optimization, which takes the dynamic electricity price into consideration. More complex fitness function is required, and the final results will be  more actionable recommendations.

\section*{Acknowledgements}
I wish to acknowledge the help provided by Lei Lu and Felix So on the great R\&D works.

I would like to express my deep gratitude to Dr. Eric Chong and Mr. Keith Cheng for offering the opportunity and their support on the works. I would also like to thank Mr. Yuelin Liang and Ms. Jimalyn Yao for their advice.

I am particularly grateful for the assistance given by Mr. Ricky Liu, Mr. Tim Lo, Ms. Rita Leung and Ms. Hetty Leung on the domain knowledge.

I would also like to extend my thanks to my Siemens colleagues, Ms. Christine Yip, Mr. Antrip Ho, and all who contributed in this work.

Assistance provided by the HKSTP colleagues Mr. Oscar Wong, Mr. Thomas Chan and Mr. Charles Leung under the umbrella of the Smart City Digital Hub was greatly appreciated.

\bibliographystyle{unsrt}


\end{document}